\documentclass[fleqn,usenatbib]{mnras}

\usepackage{newtxtext,newtxmath}

\usepackage[T1]{fontenc}
\usepackage{ae,aecompl}
\usepackage{multicol}
\usepackage{graphicx}
\usepackage{color}
\usepackage{graphicx}
\usepackage{rotating}
\usepackage{longtable}
\usepackage{lscape}
\usepackage{multirow}
\usepackage{multicol}

\newcommand{\figref}{Figure~\ref}

\newcommand{\secref}{Section~\ref}

\title[Canis Major overdensity as seen by DECam and Gaia]{A revised view of the Canis Major stellar overdensity with DECam and Gaia: new evidence of a stellar warp of blue stars}

\author[J. A. Carballo-Bello et al.]{Julio A. Carballo-Bello$^{1}$\thanks{E-mail: jcarballo@uta.cl}, David Mart\'inez-Delgado$^{2}$, Jes\'us M. Corral-Santana$^{3}$,\newauthor 
Emilio J. Alfaro$^{2}$, Camila Navarrete$^{3,4}$, A. Katherina Vivas$^{5}$ \& M\'arcio Catelan$^{6,4}$\\
$^{1}$Instituto de Alta Investigaci\'on, Universidad de Tarapac\'a, Casilla 7D, Arica, Chile\\
$^{2}$Instituto de Astrof\'isica de Andaluc\'ia, CSIC, E-18080, Granada, Spain\\
$^{3}$European Southern Observatory, Alonso de C\'ordova 3107, Casilla 19001, Santiago, Chile\\
$^{4}$Millennium Institute of Astrophysics, Santiago, Chile\\
$^{5}$Cerro Tololo Inter-American Observatory, NSF's National Optical-Infrared Astronomy Research Laboratory, Casilla 603, La Serena, Chile\\
$^{6}$Instituto de Astrof\'isica, Facultad de F\'isica, Pontificia Universidad Cat\'olica de Chile, Av. Vicu\~na Mackenna 4860, 782-0436 Macul,Santiago, Chile\\
}

\pubyear{2020}

\begin{document}
\label{firstpage}
\pagerange{\pageref{firstpage}--\pageref{lastpage}}
\maketitle

\begin{abstract}
We present DECam imaging combined with {\it Gaia} DR2 data to study the Canis Major overdensity. The presence of the so-called Blue Plume stars in a low-pollution area of the color-magnitude diagram allows us to derive the distance and proper motions of this stellar feature along the line of sight of its hypothetical core. The stellar overdensity extends on a large area of the sky at low Galactic latitudes, below the plane, and between $230^\circ < \ell < 255^\circ$. According to the orbit derived for Canis Major, it presents an on-plane rotation around the Milky Way. Moreover, additional overdensities of Blue Plume stars are found around the plane and across the Galaxy, proving that these objects are not only associated with that structure. The spatial distribution of these stars, derived using {\it Gaia} astrometric data, confirms that the detection of the Canis Major overdensity results more from the warped structure of the Milky Way disk than from the accretion of a dwarf galaxy. 
\end{abstract}


\section{Introduction}

The Galactic halo is partially the result of the continuous merging and accretion of minor 
 satellites \citep[e.g.][]{Font2011,Rodriguez-Gomez2016}. Large scale sky surveys, such as SDSS, 2MASS, DES and Gaia, have shown us the heterogeneous origin of the outer Milky Way by unveiling past accretion events \citep[e.g.][]{Helmi2018,Myeong2019,Torrealba2019}, a population of stellar tidal streams and overdensities \citep[e.g.][]{Bernard2016,Shipp2018}, and faint galaxies \citep[e.g.][]{Drlica-Wagner2015,Koposov2015,Koposov2017,Torrealba2018}. One of the main evidence of hierarchical formation in our Galaxy is the tidal stream generated by the accretion of the Sagittarius dwarf galaxy \citep{Ibata1994,Martinez-Delgado2001,Majewski2003,Belokurov2006,Koposov2012}, which is orbiting the Milky Way in almost a polar orbit. Since its discovery, a handful of new stellar overdensities, clumps and streams have been discovered in the halo and this family will keep growing as the coverage of the sky by new projects (e.g. LSST) and their completenesses increase.

The so-called Canis Major overdensity (CMa), an elliptical-shaped stellar overdensity centered at ($\ell,b$)= (240$^{\circ}$, -8$^{\circ}$), was discovered by \cite{Martin2004} by analyzing the distribution of M-giant stars below and above the Galactic plane. Such a remarkable Milky Way halo substructure is spreading over a wide area ($\sim 100$ sq deg) of the sky and seems to be located at only $<d_{\odot}> \sim 7$\,kpc \citep{Bellazzini2004,Martin2004b,Martinez-Delgado2005,Bellazzini2006a,Butler2007}.

For many years, the origin of that overdensity generated an intense debate. 
One of the hypothesis about its nature 
is that CMa is the core of a dwarf galaxy accreted at low Galactic latitudes. The first estimates, based on M-giant stars counts, suggested that CMa and the Sagittarius dwarf galaxy have a similar total mass \citep{Martin2004}. Moreover, its absolute magnitude ($M_{\rm V} \sim -14.4$) and central surface brightness ($\mu_{\rm V,0} \sim 24$\,mag\,arcsec$^{-2}$) values seemed to locate this hypothetical galaxy in the same region of the luminosity - size and $M_{\rm V} - \mu_{\rm V}$ planes where other Milky Way satellites are located \citep{Martinez-Delgado2005,Bellazzini2006a,Butler2007}. Moreover, a group of peculiar globular clusters (NGC\,1851, NGC\,1904, NGC\,2298 and NGC\,2808, among others) has been associated with CMa, what initially reinforced its definition as possible accreted galaxy within the inner Galactic halo \citep{Forbes2004,Martin2004,Forbes2010}. Those globulars present extended halos and seem to be surrounded by an unexpected stellar population, what might confirm their accreted origin \citep[see][and references therein]{Carballo-Bello2018}, as also suggested by their position in the age-metallicity distribution of Galactic globular clusters \citep[see][]{Kruijssen2019}. In addition, this family of globulars has been associated with the {\it Gaia Sausage}, a structure in the velocity space formed after the disruption of a massive galaxy \citep[see][and references therein]{Myeong2018,Forbes2020}. The scarce analyses of chemical abundances of CMa members found that some of those stars displayed abundance ratios clearly unusual for Galactic stars \citep[e.g.][]{Sbordone2005}, thus confirming the extra-Galactic nature of that halo substructure.

The excess of bright blue stars in the color-magnitud diagrams (CMD) obtained for the central regions of CMa with respect to the ones derived for the symmetric fields above the Galactic plane,  has been suggested as an additional evidence for the extra-Galactic origin of this overdensity \citep{Bellazzini2004,Martinez-Delgado2005,deJong2007}. This possibly young ($1-2$\,Gyr) stellar population, namely the Blue Plume (BP), is bright enough to be detected even in shallow wide-field photometry and is found in a section of the CMD with very low levels of contamination by fore/background stars. \cite{Dinescu2005} measured the proper motions for a sample of BP stars in direction of the CMa core ($\mu_{\rm \ell} \cos(b)=-1.47$\,mas\,yr$^{-1}$, $\mu_{\rm b} = -1.07$\,mas\,yr$^{-1}$) and derived, together with the radial velocity measured for M-giant stars in the same region \citep[][v$_{\rm r}$ = 109\,km\,s$^{-1}$]{Martin2004b}, a tentative orbit for this hypothetical dwarf galaxy. Their results show that CMa might have an in-plane rotation, similar to that of the thick-disk stars in that direction of the Galaxy but with a remarkable motion perpendicular to the plane ($W=-49$\,km\,s$^{-1}$). An alternative interpretation for the detection of BP stars along this line of sight suggests that this population, which is also detected at different longitudes across the third Galactic quadrant, might be associated with a out-of-plane spiral arm of the Milky Way instead of being part of the debris from a disrupted dwarf galaxy  \citep{Carraro2005,Moitinho2006,Carraro2008,Powell2008}. 

In fact, the Galactic origin of CMa was the alternative scenario considered since just after its discovery: the warped Milky Way disk, firstly suggested by \cite{Kerr1957}, widely studied using different tracers \citep[e.g.][]{Wouterloot1990,Lopez-Corredoira2002,Levine2006,Witham2008,Chen2019,Romero-Gomez2019} and observed in many other spiral galaxies \citep[e.g.][]{Sanchez-Saavedra1990}, might be responsible for the overpopulation of stars along the line of sight of CMa. A series of analyses of the expected and observed stellar counts \citep{Momany2004,Lopez-Corredoira2006,Momany2006,Lopez-Corredoira2007} found plausible to reproduce the observed spatial distribution of CMa stars in the third Galactic quadrant with a warped+flared disk. Two large sets of physical mechanisms have been proposed to explain the non-planarity of the Galactic disk \citep{Alfaro1996,Lopez-Corredoira2020}: (i) the former is based on the existence of a Galactic magnetic field and its possible interaction, either with the intergalactic magnetic field \citep[i.e.][]{Battaner1990}, or with the Galactic disk disturbances due to energy and momentum injection through collisions with high-velocity clouds \citep{Santillana1999,Franco2002}, dark-matter halos \citep{Bekki2009}, or satellite galaxies \citep{Hu2018}; the second group (ii) encompasses the different gravitational couplings that can occur when the angular momentum of different galactic subsystems (i.e. halo and disk) are misaligned \citep{Dubinski2009}. The first case would mainly affect gas, and only young stars, born from this material, would partially fit the gas pattern. If the stellar population, in a wide range of ages, showed the warped shape, then it would be more logical to think that the gravitational interaction between different subsystems of the Milky Way is its main driver. The latest results seem to show that the warp is present at different ranges of stellar ages which leads to think that gravitational interaction between Galactic subsystems  is the main mechanism that drives the warp formation \citep{Poggio2018}.

Moreover, the lack of an excess of RR\,Lyrae stars at the position of CMa, as previously reported for other well-studied dwarf galaxies, was pointed out in the last work on this overdensity for the last decade and discarded once again its extra-Galactic origin \citep{Mateu2009}. The absence of RR\,Lyrae stars would imply that the alleged galaxy would have an old population which is either negligible or with very unique properties. Either case implies the stellar population of CMa is at odds with what is observed in the rest of dwarf spheroidal galaxies in the Local Group and around other nearby stellar systems.

To add more complexity to the interpretation of the CMa nature, early numerical simulations of the Monoceros ring, a vast stellar halo substructure surrounding the Milky Way \citep{Newberg2002}, placed the overdensity in a projected position compatible with that of the stream \citep{Penarrubia2005}. However, this identification of CMa as the accreted progenitor galaxy of Monoceros was not conclusive and even the origin of the ring as a possible Galactic disk perturbation has recently been object of discussion \citep[e.g.][]{Xu2015}. 

In this paper, we combine new photometric data from our wide-field DECam survey of CMa with parallaxes and proper motions provided by the second {\it Gaia} data release (DR2) to map the extend of CMa and shed more light on its real nature. This dataset allows us to study the three-dimensional structure of this stellar overdensity by tracing BP stars across the area covered by our DECam survey. We also extend this study to the whole sky based on the photometry obtained by this space mission, which is crucial to confirm/discard the association of such a remarkable stellar feature with any of the components of the Milky Way.

\section{Observations and cross-match with Gaia}
\label{observations}

  \begin{figure}
     \begin{center}
      \includegraphics[width=\columnwidth]{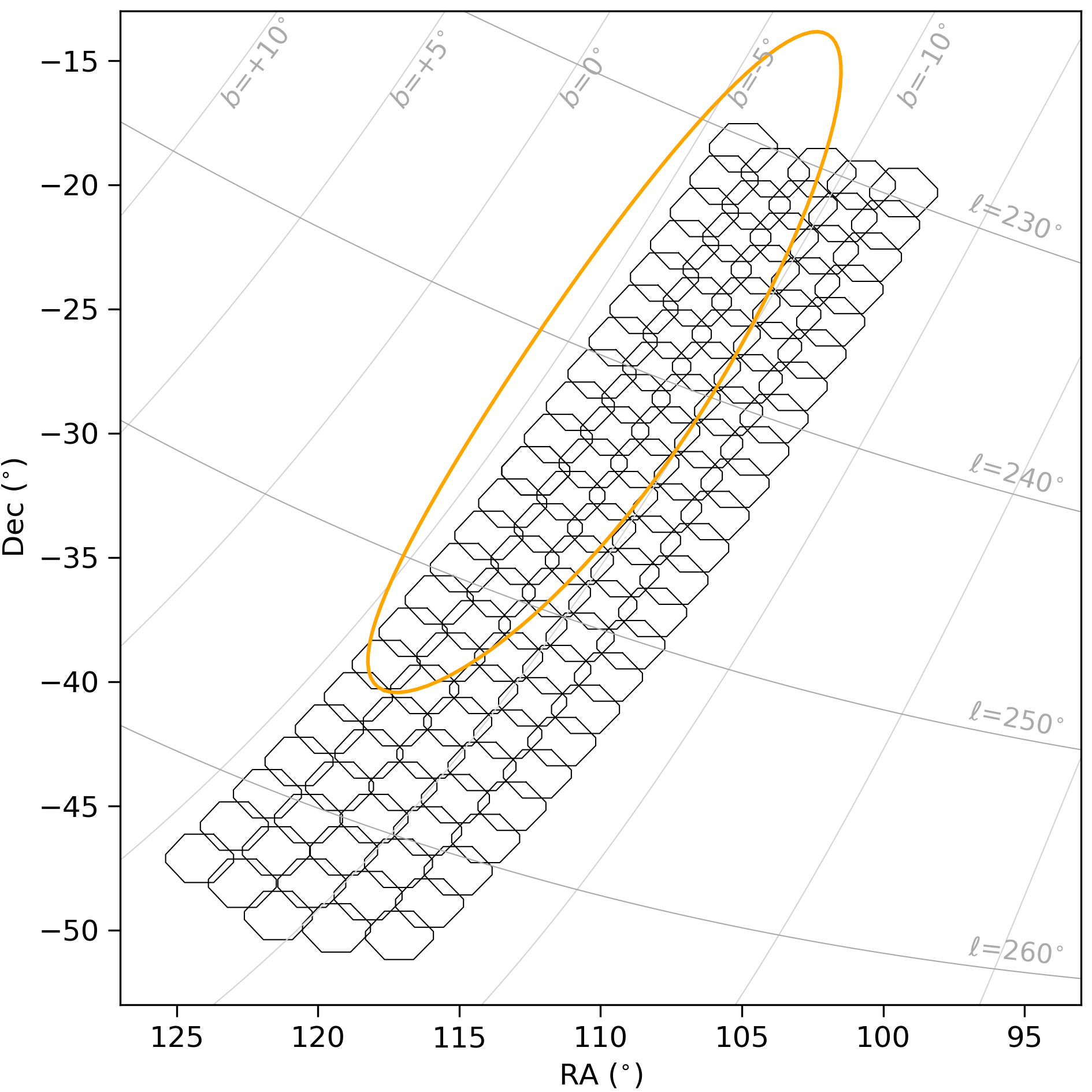}
      \caption[]{Distribution in the sky of the 115 fields observed with DECam. The solid orange line corresponds to the tentative projected position of the CMa core as described by \cite{Butler2007}.} 
\label{fields}
     \end{center}
   \end{figure}

  \begin{figure*}
     \begin{center}
      \includegraphics[scale=0.5]{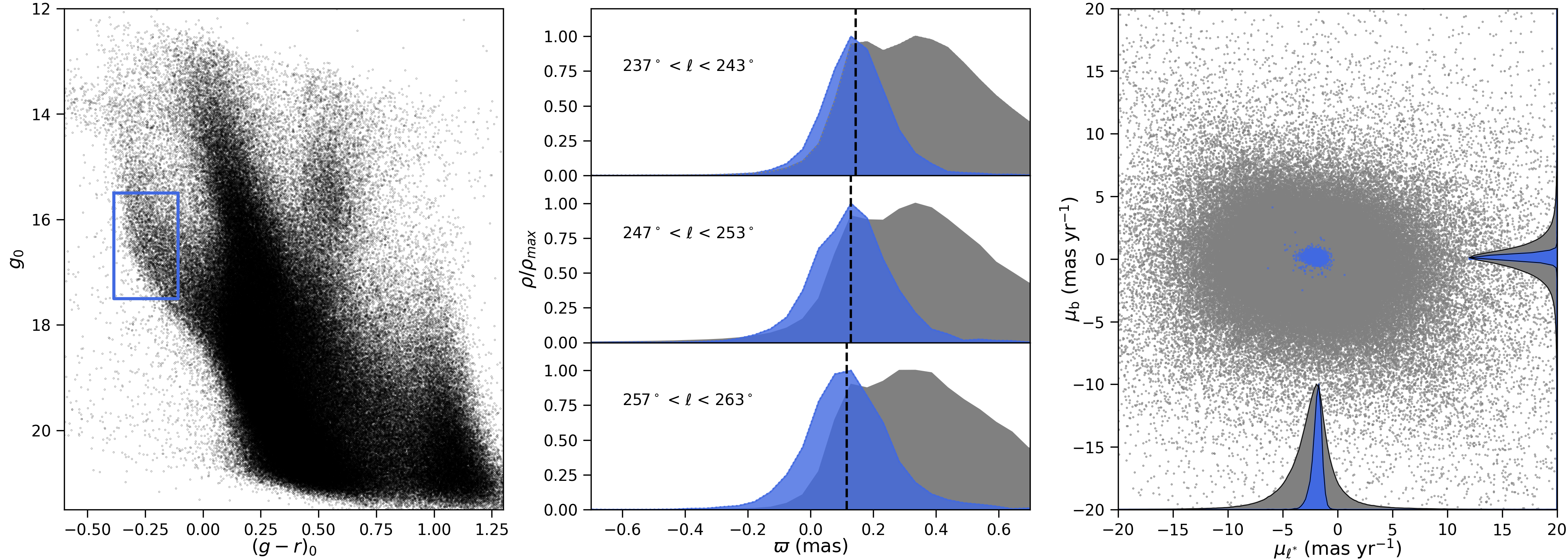}
      \caption[]{\emph{Left:} CMD corresponding to the central regions ($237^{\circ} \leq  \ell \leq  243^{\circ}$ and $-8^{\circ} \leq  b \leq  -5^{\circ}$)  of CMa in our DECam$\cap${\it Gaia}. The blue lines indicate the selection box of BP stars in the diagram. \emph{Middle}: parallax distribution of all objects (grey) and BP stars (blue) in the Galactic longitude ranges $237^{\circ} \leq \ell \leq 243^{\circ}$ (top), $247^{\circ} \leq \ell \leq 253^{\circ}$ (middle) and $257^{\circ} \leq \ell \leq 263^{\circ}$ (bottom). The dashed vertical lines indicate the central value of the Gaussian fit of the distributions of stars. \emph{Right}: distribution for all (grey) and BP (blue) stars in the proper motion space for stars with $237^{\circ} \leq \ell \leq 243^{\circ}$. In both axes are presented the scaled distributions corresponding to those populations, using the same color code.} 
\label{seleccion}
     \end{center}
   \end{figure*}

We have surveyed a region in the sky containing CMa using the Dark Energy Camera (DECam), which is mounted at the prime focus of the 4-m Blanco telescope at Cerro Tololo Inter-American Observatory (CTIO). DECam provides a 3\,deg$^{2}$ field of view with its 62 identical chips with a scale of 0.263\,arcsec\,pixel$^{-1}$ \citep{Flaugher2015}. The area to explore was confined to $230^{\circ} \leq  \ell \leq  265^{\circ}$ and $-13^{\circ} < b < -5.5^{\circ}$, as shown in \figref{fields}. The exposure time was 20\,s both for the $g$ and $r$ bands. A total of 115 DECam fields were observed in a single night (08.01.2014; proposal ID \#\,2013A-0615) with stable seeing conditions. We also observed 5 Sloan Digital Sky Survey (SDSS) fields at different
airmasses to derive the atmospheric extinction coefficients for that night and the transformations between the instrumental magnitudes and the $ugriz$ system.

The images were processed by the DECam Community Pipeline \citep{Valdes2014} and accesed via the NOAO Science Archive. The photometry was obtained from the  images with the PSF-fitting algorithm \textsc{DAOPHOT II/ALLSTAR}
\citep{Stetson1987}. The final catalog only includes stellar-shaped objects with $|sharpness| \leq  0.5$ to avoid the unavoidable pollution by background galaxies and non-stellar sources. We also used \textsc{DAOPHOT\,II} to include in our images synthetic stars  with magnitudes in the range $17 \leq  g,r \leq  25$ and $0 \leq  g-r \leq  1.5$ in order to estimate the completeness of our photometry. We applied our photometry pipeline with these altered images and the limiting magnitude was set at that of the synthetic stars that are only recovered in 50$\%$ of the cases. For $g$ and $r$ bands, the limiting magnitude in our catalogs are 20\,mag and 20.5\,mag, respectively. Extinction coefficients have been derived\footnote{\url{https://github.com/kbarbary/sfdmap}} from \cite{Schlafly2011} dust maps and we adopted the multiplicative coefficients estimated by the Dark Energy Survey collaboration \citep[$R_{\rm g}=3.186$ and $R_{\rm r}=2.140$;][]{Abbott2018}.

The European Space Agency (ESA) mission {\it Gaia} will provide precise positions, kinematics and stellar parameters for more than one billion stars aiming to understand the origin and evolution of our own Galaxy \citep{Gaia2016}. So far, the second data release \citep{Gaia2018a} has provided the five-parameter astrometric solution (positions, proper motions and parallaxes), making it possible to explore the structure of the Milky Way with unprecedented detail. In order to include this precious information in our final photometric catalog, we crossed-match our DECam results with the {\it Gaia} DR2 database. 

Among the available information for each source, we ensure a good quality photometry and astrometry by using the $BP/RP$ excess factor, which measures the excess of flux in the $G_{\rm BP}$ and $G_{\rm RP}$ integrated photometry with respect to the $G$ band, and the number of visibility periods used in the astrometric solution, i.e., the number of distinct observation epochs. For this work, only those sources in {\it Gaia} with \textsc{phot\_bp\_rp\_excess\_factor} $\leq$  1.5 and \textsc{visibility\_periods\_used} $\geq$ 5 were considered. We also adopted the formalism of the renormalized unit weight error \citep[RUWE;][]{Lindegren2018b} and we assumed that only objects with RUWE $\leq$ 1.4 have an acceptable astrometry. Despite the slightly shallower photometry provided by that mission, only an average of $5\%$ of objects per field lack {\it Gaia} counterpart satisfying the conditions described above. Therefore, our DECam$\cap${\it Gaia} dataset is almost identical to our initial photometry, with a final sample of stars of 11,663,496 stars. We have used the {\it Gaia} extinction coefficients provided by \cite{Gaia2018a} and the photometric errors were computed using $\sigma_{\rm mag}^{2} = (1.086\sigma_{\rm flux}/{\rm flux})^{2} + (\sigma_{\rm zp})^{2}$, where $\sigma_{\rm flux}$ and $\sigma_{\rm zp}$ are the error in the flux and photometric zeropoint, respectively \citep{Evans2018}. 

\section{Results and discussion}

\subsection{Tracing the Canis Major structure with Blue Plume stars in the DECam photometry}

BP stars have been considered the best stellar tracers of the presence of CMa in the third Galactic quadrant. Indeed, in the CMD contained in \figref{seleccion} corresponding to the center of the overdensity \citep[$\ell \sim 240^{\circ}$ , according to the red-clump (RC) stars distribution obtained by][]{Bellazzini2006a}, a feature composed of blue stars is observed in the ranges $(g-r)_{0} < 0$ and $15 < g_{0} < 18$. BP stars are very bright and significant bluer than the rest of foreground and background Milky Way population. This enables to use them as excellent tracers of the possible CMa population across the full sky, being also ideal targets for follow-up spectroscopy studies (even with intermediate-size telescopes) to derive their kinematics and chemical abundances. In order to obtain a sub-sample of CMa star candidates based on the proper motions and parallaxes provided by {\it Gaia}, we have selected the BP stars with $-0.4 \leq  (g_-r)_{\rm 0} \leq  -0.1$, $15.5 \leq  g_{\rm 0} \leq  17.5$, $237^{\circ} \leq  \ell \leq  243^{\circ}$ and $-8^{\circ} \leq  b \leq  -5^{\circ}$. Since hot subdwarf stars are the only possible polluters in this region of the CMD, we have checked that none of our BP stars are found in the \cite{Geier2019} catalogue. The selected section of the CMD is indicated in the left panel in \figref{seleccion}.

\subsubsection{Parallaxes and distance}

Parallax and distance are two related variables but they do not follow a linear
function. Thus, inferring the distance from the angular parallax, either for
individual stars or for coherent and compact stellar systems is not an easy task.
With the publication of {\it Gaia}'s data this topic has received a lot of attention, and
one can find numerous works in recent literature where various methods are
proposed to estimate the distance from parallax and its error’s distribution \citep[i.e.][among others]{Lindegren2018,Luri2018,Bailer-Jones2018,Chen2018,Maiz-Apellaniz2019}.

The parallaxes catalogued in {\it Gaia} DR2 seem to be too small. The analysis of
the zero-point correction provides different results depending on the type of
objects studied \citep{Lindegren2018,Maiz-Apellaniz2019,Chan2020}, their location within the Galaxy \citep{Arenou2018}, the spatial coherence of the sample \citep{Khan2019}, the apparent magnitudes \citep{Arenou2018}, or their photometric colors \citep{Chen2018,Chan2020}. Most of these
corrections are between 0.03 and 0.05 mas. We have corrected the listed
parallaxes by a constant zero point of 0.04 mas, which is a representative value
of the proposed zero-points distribution and coincides with the one suggested by
\citet{Maiz-Apellaniz2019} for the OB stars in the GOSSS catalogue. The same happens with the parallax error in {\it Gaia} DR2 that only reports the internal precision. A correction for the catalogued errors is given by the
equation $\sigma^{2}_{total} = K\,\sigma^{2}_{int} + \sigma^{2}_{ext}$,
where $K =$ 1.08 and $\sigma _{ext} = 0.043$\,mas for stars with G $\geq$ 13\,mag, as recommended by \cite{Lindegren2018}.

We have obtained the parallax distribution for all the stars in the field brighter than $g_{\rm 0} = 18$ and the BP stars, weighted by their error ($w_{i} = 1 / \sigma_{\rm total_{i}}^{2}$), and using a bin size of $\delta_{\varpi}  = 0.05$\,mas. Although for single objects negative parallaxes make no physical sense, their removal from the analysis of a stellar system would bias the parallax distribution yielding erroneous estimates of its statistical moments \citep[e.g. see][]{Luri2018}. Selection by parallax internal precision could also introduce some bias hard to detect and remove in most cases. We thus have made use of the complete photometric sample, just pruned from Gaia bad quality data choosing those objects matching the RUWE condition described in \secref{observations}. 

The resulting distributions are shown in the upper middle panel of \figref{seleccion}. While most of the field stars are concentrated in the $\varpi > 0.2$\,mas, peaking at $\varpi \sim 0.35$\,mas, the distribution for BP stars shows a fairly symmetrical distribution where mode, median, and mean are centred around 0.14\,mas, thus suggesting the presence of a well differentiated substructure along this line of sight. Two-sample Kolmogorow-Smirnov and Wilcoxon tests show that both samples represent different statistical populations in term of parallaxes, at a confidence level of 99$\%$. The error in the weighted mean, for our selected BP stars, can be approached by $\epsilon = \sqrt{1/\sum w_{i}}$ \citep{Chen2018}. That error is 0.00008\,mas, too low to be representative of the central parallax total uncertainty. The {\it Gaia} parallax zero-point correction applied here has taken into account only the blue color of these objects and choosing as a representative value the one given by \cite{Maiz-Apellaniz2019} for OB stars. However, other authors propose a value of 0.03\,mas for the same spectral-type stars. These approaches lead us to suggest that a more realistic estimate of the uncertainty in the obtention of the central parallax for CMa BP stars is not under 0.01\,mas. We conclude that the population traced by the BP stars is located at an average distance of 7.1\,$_{-0.4}^{+0.6}$\,kpc, which is in good agreement with the values previously reported \citep[$<d_{\odot}> \sim 7$\,kpc, ][]{Martin2004,Martinez-Delgado2005,Bellazzini2006a,Butler2007}. We repeat this procedure for two additional Galactic longitude ranges, which show similar distributions, as shown in the central panel in \figref{seleccion} and derived the following heliocentric distances: 7.8\,$_{-0.6}^{+0.7}$\,kpc ($\ell=250^{\circ}$) and 9.0\,$_{-0.7}^{+1.0}$\,kpc ($\ell=260^{\circ}$). Therefore, the substructure traced by BP stars is observed throughout the full region observed with DECam, $230^{\circ} \leq  \ell \leq  265^{\circ}$, and it is confined in the $6.7 < d_{\odot}{\rm [kpc]} < 10.0$ section of our DECam$\cap${\it Gaia} catalog.

\subsubsection{Proper motions}
\label{proper}

Once we have analyzed the spatial distribution of BP stars along the line of sight towards the central region of CMa, we proceed to derive mean proper motions for the bulk of CMa stars by using the same blue stars as reference.  Proper motions have been transformed to the Galactic system using the equations found in \cite{Poleski2013}. As shown in the right panel in \figref{seleccion}, BP stars are concentrated in the ($\mu_{\rm \ell}\cos(b), \mu_{\rm b}$) plane around $<\mu_{\rm \ell} \cos(b)> = -1.78 \pm 0.38$\,mas\,yr$^{-1}$ and $<\mu_{\rm b}> = 0.13 \pm 0.24$\,mas\,yr$^{-1}$, where the quoted error means for the standard deviation of the distributions. The mean values obtained for the BP stars are slightly shifted with respect to the ones derived for the complete sample, and with a smaller standard deviation, thus proving that BP stars in the CMa field are concentrated on both parallax and proper motions variables and represent a well-defined stellar population in the {\it Gaia} astrometric space.  

In order to better draw the contours of this population in the 3D space defined by proper motions and parallax, we have obtained the Spectrum of Kinematic Grouping (SKG) for the BP stars. This method was developed by \cite{Alfaro2016} to analyse spatial groupings of stellar radial velocity in star-forming regions, and has been applied to the search for kinematic substructures in selected regions, including NGC\,2264 \citep{Gonzalez2017}, Cyg\,OB1 \citep{Costado2017}, and Monoceros \citep{Costado2018}. This method is based on that developed by \cite{Allison2009} with the contributions of \cite{Maschberger2011} to analyse the mass segregation in stellar clusters and star-forming regions. Basically, it consists in sorting the variable to be analysed, parallax in our case, and splitting the sample into Nint intervals of the same number of Ndat data, which are initially taken as Ndat = $\sqrt{\rm Ntotal}$. For each parallax interval, the median of the Minimum Spanning Tree (MST) edges is calculated and compared with the same value obtained for a random selection of Ndat points extracted from the sample. If the median of the random selection is significantly greater than that of the analysed interval, we say that this parallax presents a grouping in the Vector- Point Diagram (VPD). The ratio between the median of the MST edges of the random sample and that of the corresponding interval ($\Lambda$) has to be greater than 1 (at least with a 2 $\sigma$ confidence interval) to consider that the representative parallax of that interval is segregated. The specifics of the method and its application can be seen in more detail in \cite{Alfaro2016} and \cite{Alfaro2018}.

  \begin{figure}
     \begin{center}
      \includegraphics[scale=0.5]{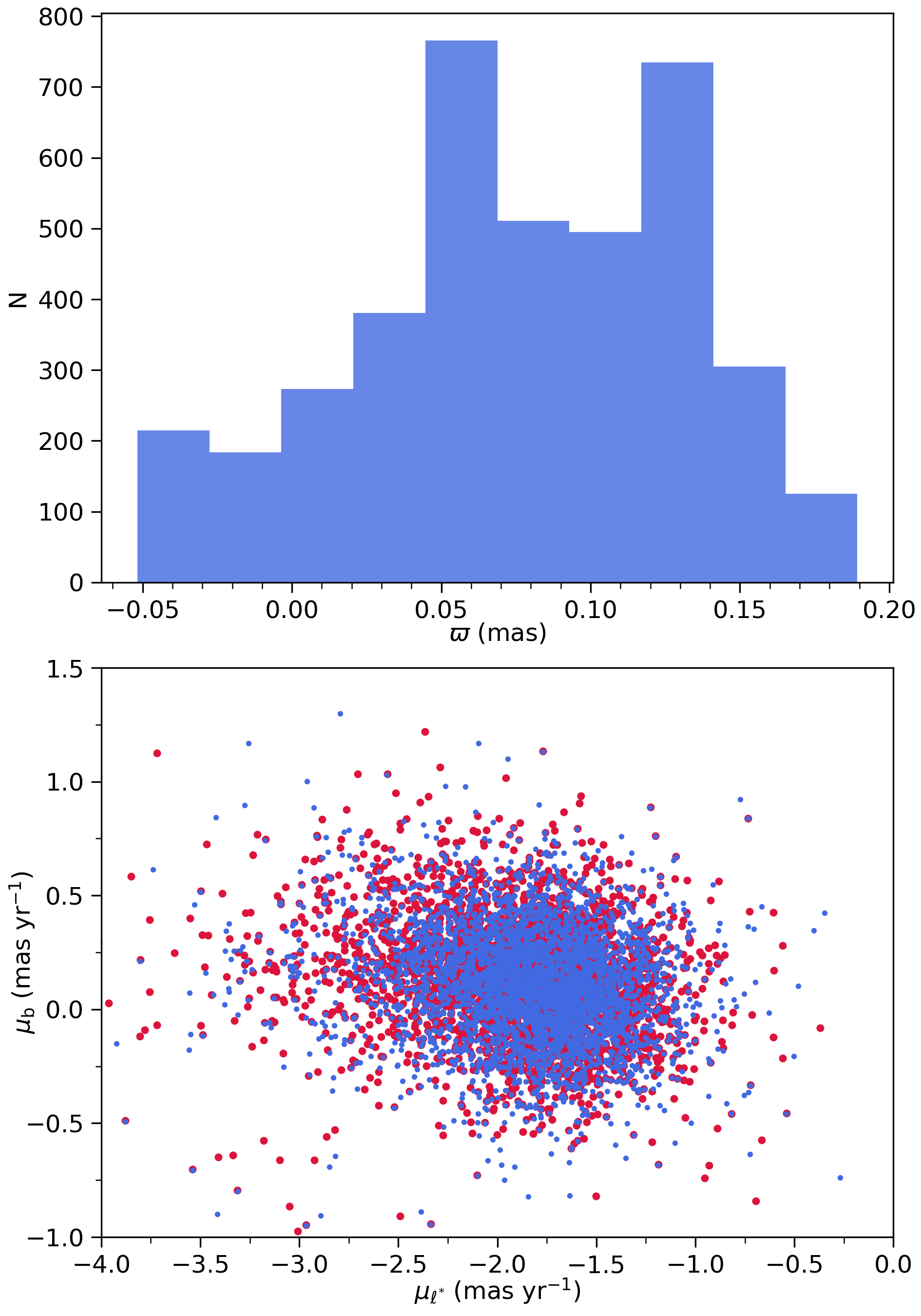}
      \caption[]{\emph{Top:} Parallax distribution for the $\sim$ 4000 stars resulting from the SKG analysis.  \emph{Bottom:} distribution in the proper motion space for the photometrically (red) and SKG selected (blue) BP stars.} 
\label{plotskg}
     \end{center}
   \end{figure}

For the SKG analysis of the $\sim 15000$ BP stars in $237^{\circ} \leq \ell \leq 243^{\circ}$ and $-8^{\circ} \leq b \leq -5^{\circ}$, we have chosen bins of 120 objects with an overlap of 30 between two consecutive intervals. If we select the bins with $\Lambda - 2\,\sigma$ > 1 we find that the stars in those bins ($\sim$ 4000) show a bimodal distribution with two maxima peaking at $\varpi \sim$ 0.12 and 0.06\,mas (see upper panel in \figref{plotskg}). The VPD of the SKG selected stars is shown in the lower panel in \figref{plotskg}, where a non-circular structure with a greater elongation in the $\mu_{\rm \ell} \cos(b)$ axis is manifested. It should be noted that SKG analysis points to the existence of  concentrations in the 3D space formed by the proper motions and parallaxes, but it does not perform a membership analysis, so this diagram may be contaminated by a few stars located very far from the central values of the 3D distribution.  

SKG analysis of the selected BP stars modifies the parallax distribution obtained for the initial sample. The main difference is its bimodal character and that the peaks’ parallaxes are lower than the weighted average value found for the total sample of BPs. There appears to be two concentrations of BP stars with coherent motions that are further apart from the Sun than the previous estimate for the distribution of BPs selected only by their position on the CMD. Taking into account the possible biases that can be introduced into the transformation of parallaxes to distances, we will adopt here $d_{\odot} \sim $8 and 16\,kpc as the most probable value for the distances of the BP concentrations, with a total error of not less than 2 kpc, based on the position of the maximums and their widths. The most distant of these components might be associated with the farthest stars detected by \cite{Carraro2017} during a study of BP stars in the neighbourhood of the sky area covered by our DECam$\cap${\it Gaia} data. It is difficult to realistically assess the global uncertainty (precision and bias) of these estimates, but our analysis shows a clear fact, not to be missed on how to perform the transformation between parallaxes and distances: there are two localized concentrations at different parallaxes (distances), both farther from the Sun than the 7.1\,$_{-0.4} ^{+0.6} $\, kpc determined  as the most representative value for the  whole sample of BPs.

\subsubsection{Spatial distribution}

  \begin{figure}
     \begin{center}
      \includegraphics[scale=0.5]{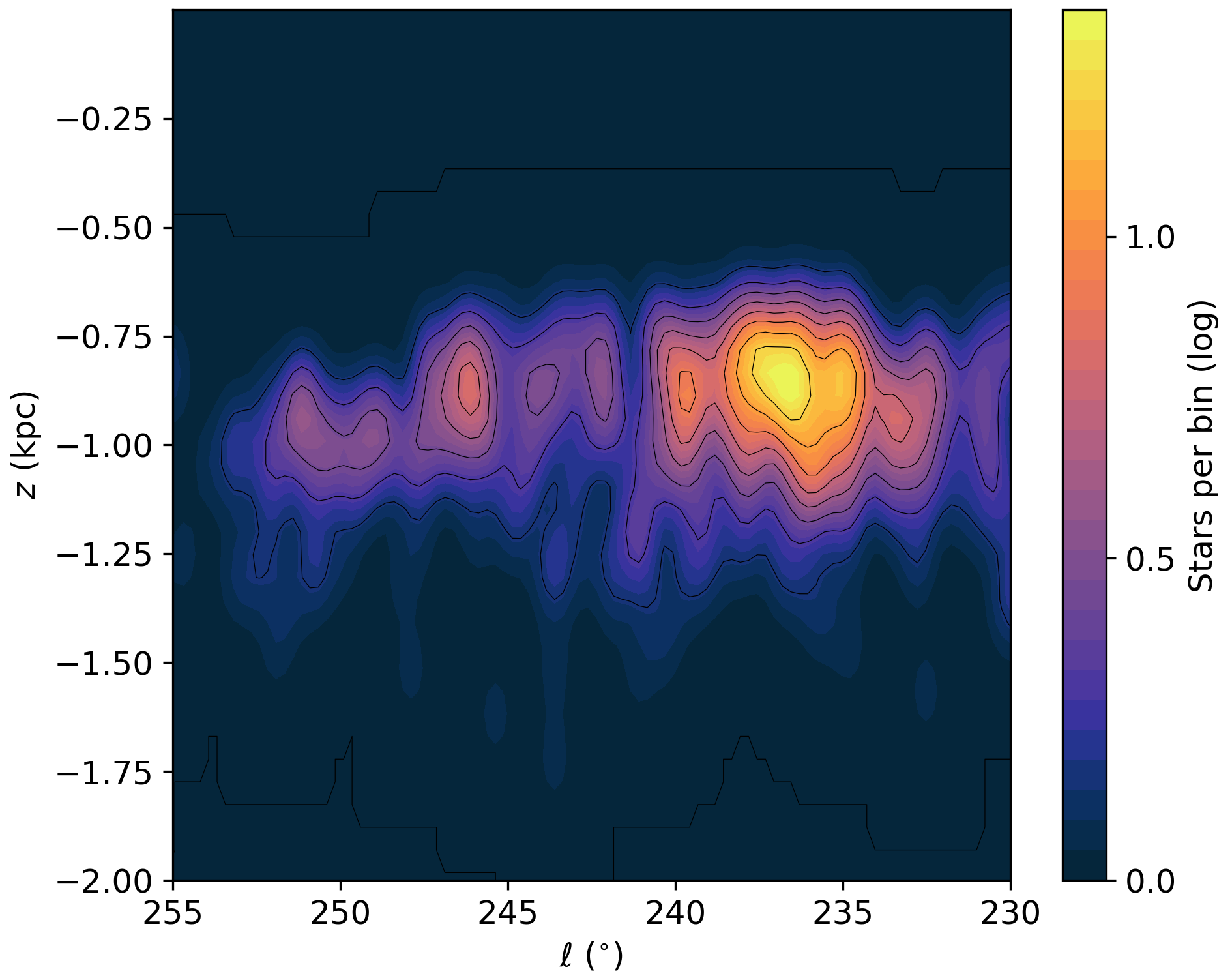}
      \caption[]{Density map generated with all the BP stars with $15.5 \leq  g_{\rm 0} \leq  17.5$, $-0.4 \leq  (g-r)_{\rm 0} \leq  -0.1$ and $E(B-V) \leq  0.3$ in our DECam$\cap${\it Gaia}.We only considered those stars satisfying the criteria in the proper motion space derived in this work for CMa and assumed a heliocentric distance of d$_{\odot}$ = 8\,kpc for the complete sample.} 
\label{mapadecam}
     \end{center}
   \end{figure}

If CMa is composed of the remnants of an accreted dwarf galaxy, the young and bright BP stars would be clustered towards its densest and central regions \citep[see e.g. Phoenix dwarf and Antlia dwarf irregular galaxies;][]{Aparicio1997,MartinezDelgado1999}. The density map shown in \figref{mapadecam} was built by counting BP stars with $E(B-V) \leq  0.3$ and matching the selection criteria in proper motions derived above, for which we assumed a distance of $d_{\odot}$ = 8\,kpc. The CMa overdensity, more specifically its hypothetical core, spans $\sim 25^{\circ}$ in Galactic longitude  as an asymmetrical elongated substructure with its densest section in $235^{\circ} < \ell < 240^{\circ}$ and confined to the $ -1 < z [\rm kpc] < -0.5$ section of the Galaxy. Although it would be also desirable to study the complete shape of the stellar overdensity by considering MS stars, the larger errors in the proper motions provided by {\it Gaia} DR2 for sources fainter than $G \sim 18$ \citep[see e.g.][]{Evans2018} make it difficult to properly distinguish between CMa and Galactic members in the faint end of our photometry. Even so, the spatial distribution traced by the BP stars identified in our DECam$\cap${\it Gaia} survey, is compatible with the presence of a stellar system with a small dispersion along this line of sight.

\subsubsection {What are the BP stars?}

With the purpose of exploring the nature of the BP stars, we have chosen the brightest objects from our sample with $g$  < 16.5\,mag. With this criterion, we try to ensure that these objects have photometric measurements in different public catalogues. To determine their physical properties, we resorted to the Virtual Observatory (VO) tool called Virtual Observatory SED Analyzer \citep[VOSA:][]{Bayo2008} that allows the obtaining of the photometric data associated with the VOSA databases and its comparison with different stellar libraries. For this particular case, we have selected the \textsc{BT-COND}, Kurucz, \textsc{TLUSTY} and Koester sets of stellar models \citep{Kurucz1993,Hubeny1995,Allard2012,Koester2010}. The first two encompasses a broad set of physical parameters, where \textsc{BT-COND} includes the coldest objects. \textsc{TLUSTY} models expand to highest temperatures and Koester models stand for white dwarfs. We leave the reddening parameter free, with $A_{\rm V}$ between 0 and 2\,mag. 

  \begin{figure}
     \begin{center}
      \includegraphics[scale=0.5]{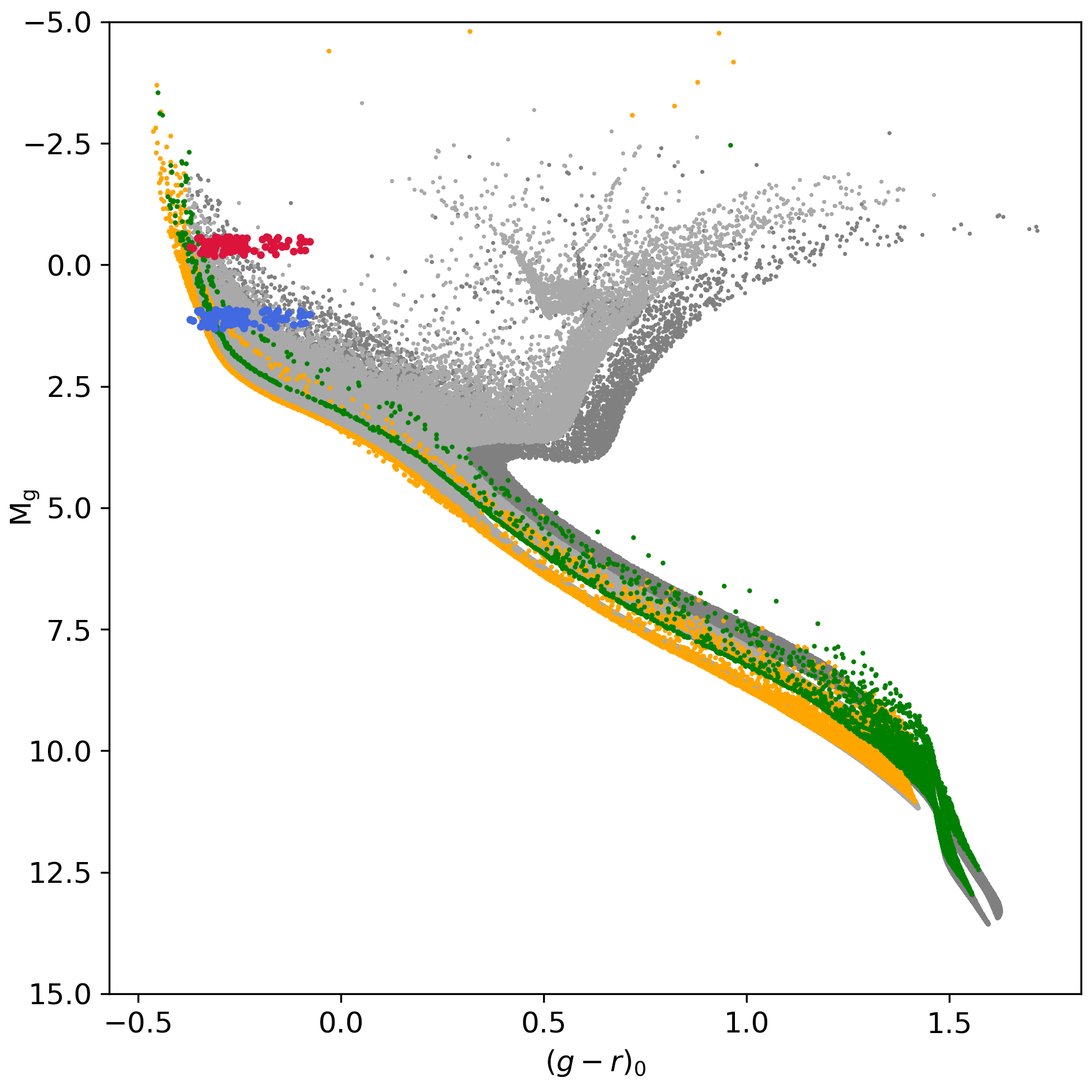}
      \caption[]{Synthetic CMDs generated using the \textsc{BaSTI} stellar evolution models \citep[][and references therein]{Hidalgo2018}  for two populations with metallicities [Fe / H] = -0.4 (light grey) and [Fe / H]= -0.9 (dark grey), and covering an age range from 20\,Myr to 10\,Gyr. Yellow and green dots indicate objects with age $ < 100$\,Myr for both models. The 92 stars with best VOSA fitting are placed in the diagram assuming the heliocentric distances to the spatial concentrations of BP stars found in our analysis at $d_{\odot} \sim 8$\,kpc (blue) and $\sim 16$\,kpc (red). Dereddened magnitudes in our catalogue  have been used for this plot. }
\label{sintetico}
     \end{center}
   \end{figure}

Of the 176 selected stars, VOSA finds a solution for 175 objects. From them, we have chosen those with $\chi^{2}$ values ​​less than 3 and a number of fitted photometric bands greater than or equal to 12. For the 92 objects that meet these conditions, the following mean values ​​are obtained:  $<T_{\rm eff}> = 9440$\,K, $ <log (g)> = 3.6$ and $<[Fe / H]> = -1.6$. The ranges of values ​​obtained expand from 7400 to 15000\,K for $T_{\rm eff}$, between 2.5 and 4.5 for $log(g)$ and -4 and 0.5 for metallicity. The metallicity value is the one with the highest degree of uncertainty, especially if we move towards high temperatures and we have no information on the ultraviolet bands. In other words, if we leave the metallicity aside, the BP stars are subgiant and dwarf objects in a temperature range perfectly compatible with their position in the CMD at the distances estimated for the two stellar concentrations detected in this work  (see \figref{sintetico}).

\subsubsection{Orbit}
\label{orbita}

We use the \textsc{galpy} package \citep{Bovy2014} to compute here a tentative orbit for a sample of 100 likely CMa BP stars using the mean proper motions and heliocentric distance ranges derived in this section. The distance of the Sun to the Galactic center and its circular velocity are set to $R_{\odot} = 8$\,kpc and $V_{\odot} = 240$\,km\,s$^{-1}$, respectively \citep{Reid2014}. The errors in our input parameters are high due the distribution observed in the proper motion space along the sky area contained in the DECam$\cap${\it Gaia} survey and the intrinsic errors in our distances to the BP stars. Therefore, although the computed orbit might be enough to explore the likely in-plane rotation of this halo substructure, it is not simple to derive a well-constrained orbit of CMa with this dataset.

  \begin{figure}
     \begin{center}
      \includegraphics[scale=0.5]{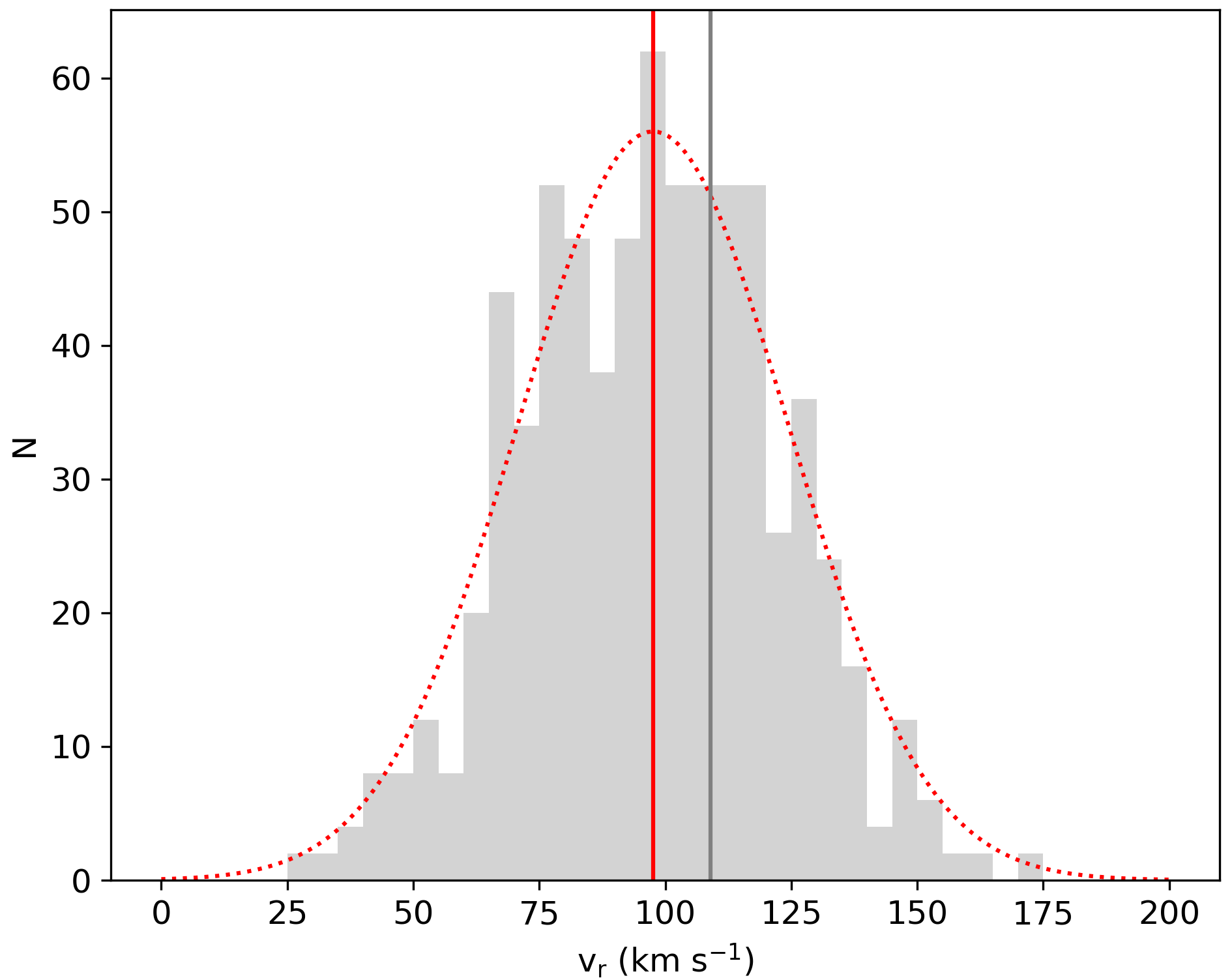}
      \caption[]{Radial velocity distribution for stars along the line of sight of the hypothetical core of CMa ($\ell, b) = (240^{\circ}, -7^{\circ})$, in the same heliocentric distance and proper motions ranges than the selected BP stars in this sky area. The dashed red line corresponds to the gaussian fitted to the distribution with a central value of $<v_{\rm r \,\, Gaia}> = 97 \pm 27$\,km\,s$^{-1}$ (vertical red line). The solid grey line indicates the mean radial velocity of CMa derived by \cite{Martin2004b}.}
\label{velocidades}
     \end{center}
   \end{figure}

  \begin{figure*}
     \begin{center}
      \includegraphics[scale=0.5]{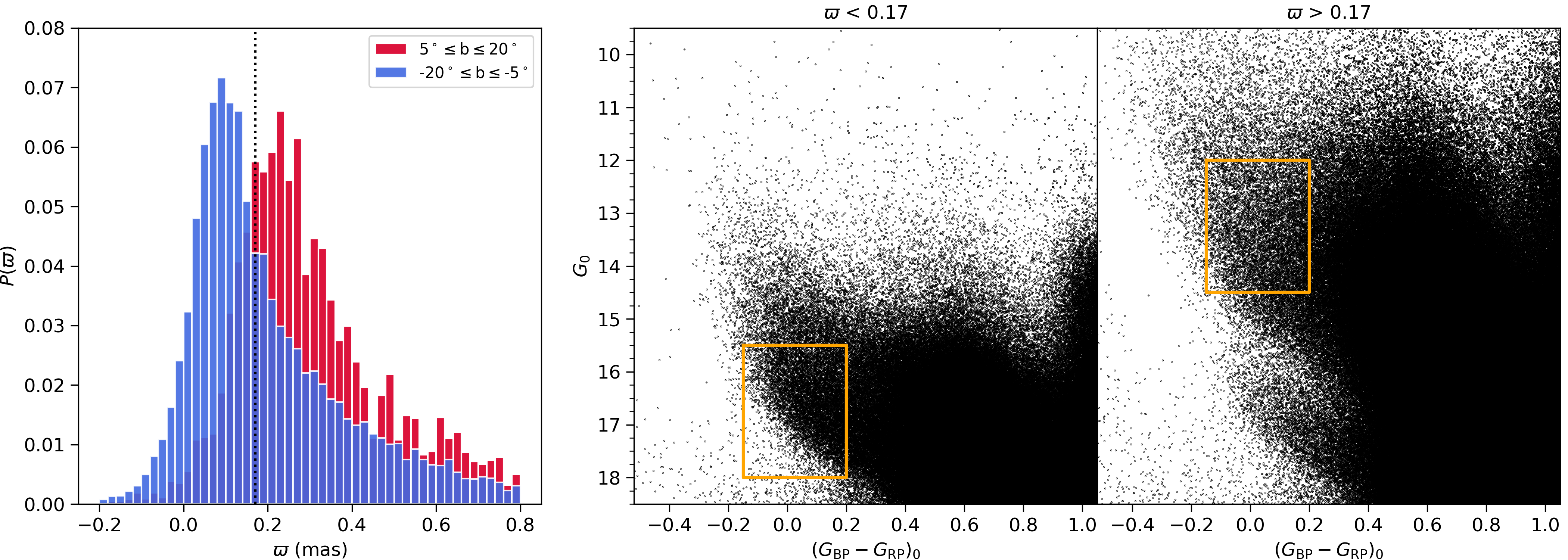}
      \caption[]{ \emph{Left}: parallax probability distribution for stars with $-0.15 \leq  (G_{\rm BP}-G_{\rm RP})_{\rm 0} \leq  0.15$ and $12 \leq  G_{\rm 0} \leq  18$. The distributions corresponding to stars in $240^{\circ} \leq  \ell \leq 245^{\circ}$, above $b=5^{\circ}$ and below $b=-5^{\circ}$ are represented in red and blue, respectively. The dotted vertical line correspond to an arbitrary separation between two populations at $\varpi \sim 0.17$\,mas. \emph{Middle} and \emph{right} panels show the CMDs correponding to all stars with $E(B-V) < 0.3$ in the region defined by  $240^{\circ} \leq  \ell \leq 245^{\circ}$, $|b| \geq 5^{\circ}$ with $\varpi \leq 0.17$ and $> 0.17$, respectively. The orange solid lines represent the selection boxes used to generate the faint and bright samples of tentative BP stars.} 
\label{comparacmds}
     \end{center}
   \end{figure*}

There is not radial velocities available for BP stars along this line of sight in {\it Gaia} DR2 yet, where only brighter and redder objects have measurements. For this reason, we retrieve all the objects with radial velocities available and satisfying the criteria derived above both in parallax and proper motions. The distribution of radial velocities for stars within a circle with radius  $r = 10^{\circ}$ centered in $(\ell,b) = (240^{\circ}, -7^{\circ})$ is shown in \figref{velocidades}. A mean radial velocity of $<v_{\rm r \,\, Gaia}> = 97 \pm 27$\,km\,s$^{-1}$ is obtained from the Gaussian fit of the distribution, which is similar to the value derived by \cite{Martin2004b} for a sample of M-giant stars likely associated with CMa and set at $<v_{\rm r \,\, M2004}> = 109$\,km\,s$^{-1}$, and with the mean velocity derived for a few tentative BP stars towards the open cluster Tombaugh\,1 \citep{Carraro2017}. The resulting orbit for CMa moving with $<v_{\rm r \,\, Gaia}>$ in a Milky Way-like potential (MWPotential2014 in \textsc{galpy}) has $<|z|_{\rm max}> \sim 1$\,kpc, while when we compute it by assuming $<v_{\rm r \,\, M2004}>$ we obtain $<|z|_{\rm max}> \sim 0.9$\,kpc. Therefore, the orbit of CMa as an hypothetical accreted dwarf galaxy is confined to the $|z| < 1$\,kpc section of the Galaxy, in contrast to the $|z|_{\rm max} = 2$\,kpc derived by  \cite{Dinescu2005}. As for the motion of CMa perpendicular to the Galactic disk, we derive $<|v_{\rm z}|_{\rm max}> \sim$ 27\,km\,s$^{-1}$ using both radial velocities, which is a value fully compatible with the on-plane rotation of the rest of Milky Way disk stars.

In summary, our analysis of the DECam$\cap${\it Gaia} data confirms that CMa is a vast and coherent stellar substructure located in the inner halo in the range $230^{\circ} < \ell < 255^{\circ}$ with a denser region containing blue and (possibly) young stars around $\ell = 237^{\circ}$. However, its orbit, based on its mean proper motions, is still compatible with that of the rest of Milky Way disk stars. This means that, with our photometric survey and the kinematical data available in the literature and {\it Gaia} DR2, CMa is confined to the plane and it seems unlikely a different origin for this overdensity from that of the rest of the stars along this line of sight in the Galaxy.

\subsection{All-sky distribution of BP stars: the spiral structure of the Milky Way}

  \begin{figure*}
     \begin{center}
      \includegraphics[scale=0.5]{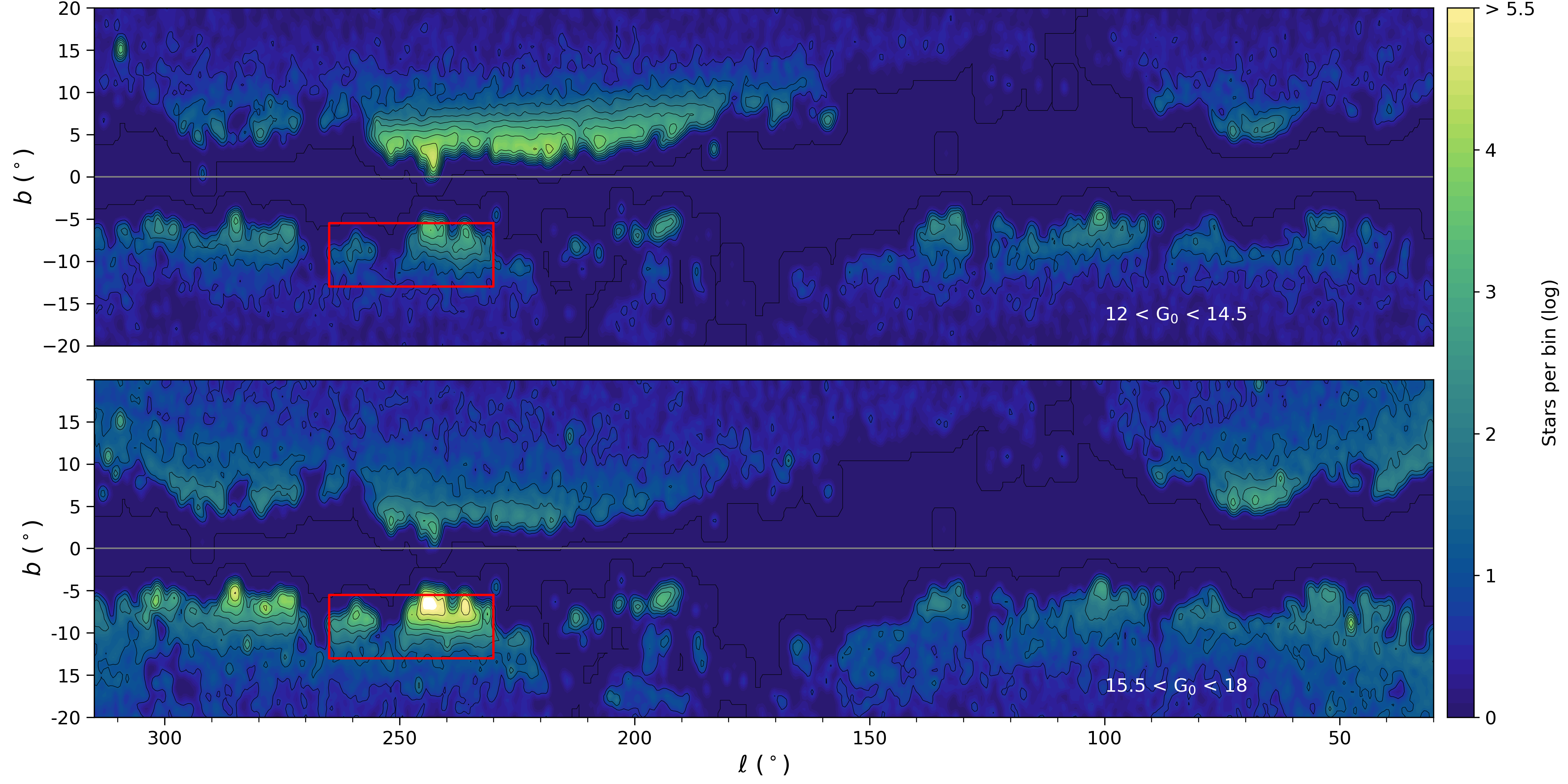}
      \caption[]{Density maps in Galactic coordinates generated with all the tentative BP stars in {\it Gaia} DR2 with colors $-0.15 \leq  {(G_{\rm BP}-G_{\rm RP})}_{\rm 0} \leq  0.15$, and satisfying $12 \leq  G_{\rm 0} \leq  14.5$ (bright sample, up) and $15.5 \leq  G_{\rm 0} \leq  18$ (faint sample, down), with $E(B-V) \leq  0.3$. Only those objects with RUWE $\leq$ 1.4 are considered. The high-density region in $220^{\circ} \leq \ell \leq 300^{\circ}$ and  $b < 0^{\circ}$ in the lower panel might be associated with the structure containing CMa. The red box indicates the sky area covered by our DECam$\cap${\it Gaia} survey.}
\label{mapaBPs1}
     \end{center}
   \end{figure*}

Is CMa part of a larger Galactic component?   The structure of the outer edge of spiral disks is still a matter of debate \citep[][and references therein]{Bland-Hawthorn2016}. In particular, for the Milky Way, some authors invoke the truncation of the disk at a Galactocentric radius of between 10 and 16 kpc \citep[e.g.][]{Minniti2011}, while others provide evidence that both the stellar and the gas components, continuously distribute up to 20 or 30 kpc from the Galactic center \citep{Lopez-Corredoira2018}. In the direction of our DECam photometric survey, the situation is more complicated because this area has shown some peculiarities, including  \emph{i)} lower levels of interstellar extinction; \emph{ii)} it coincides with the direction of the maximum southern warping of the gas and the stellar component \citep[e.g.][]{Chen2019}; \emph{iii)} there is strong evidence of a flare (an increase in the height scale with the Galactocentric distance) in the distribution of stars and gas \citep{Li2019}; \emph{iv)} the distribution of atomic and molecular gas shows gaps at different velocity ranges \citep[see pioneer work by][]{Lindblad1967}; and \emph{v)} the presence of the Monoceros halo substructure, whose nature is still unclear, has been widely confirmed \citep[e.g.][]{Morganson2016}.  

To shed more light on the origin of CMa, we proceed now to inspect for the first time its full extension in the sky making use of the ($G,~G_{\rm BP},~G_{\rm RP}$) photometry provided by {\it Gaia}.  We have retrieved  all the information for the sources from the {\it Gaia} DR2  with \textsc{phot\_bp\_rp\_excess\_factor} $\leq  1.5$ and \textsc{visibility\_periods\_used} $\geq$ 5 for the  section of the Milky Way with $|b| \leq  20^{\circ}$ and, to avoid the crowded regions of the Galaxy, $30^{\circ} \leq  \ell \leq  330^{\circ}$.

A first step is to investigate whether BP stars are only found in the CMa region or if they are also present in the northern Galactic hemisphere. We focus on the blue and bright objects  found in the $240^{\circ} \leq  \ell \leq 245^{\circ}$, $|b| \geq 5^{\circ}$ section of the Galaxy with $-0.15 \leq  (G_{\rm BP}-G_{\rm RP})_{\rm 0} \leq  0.15$, $12 \leq  G_{\rm 0} \leq  18$, and $E(B-V) \leq 0.3$. The range of $ G_{\rm 0}$ magnitudes considered allows the inclusion of brighter (and likely closer) BP stars. The left panel in \figref{comparacmds} shows the parallax probability distributions for these tentative BP stars in both Galactic hemispheres, separately. Once again, the BP stars associated with CMa (without SKG analysis) peak at $d_{\odot} \sim 7$\,kpc, while the objects at positive Galactic latitudes seem to peak at $d_{\odot} \sim 4$\,kpc. When we arbitrarily split the CMD corresponding to this region into two components with parallaxes greater than or lesser than $\varpi \sim $ 0.17\,mas, we obtained the diagrams contained in the middle and right panels in \figref{comparacmds}. We recover the characteristic BP classically associated with CMa but also a brighter/closer feature, composed of stars with $G_{\rm 0} \leq 15$. This result confirms  that the so-called BP stars also represent a prominent population in the northern hemisphere, but it might have been missed by previous photometric studies of the area because of their proximity to us,  thus yielding in brighter magnitudes and a lower density of objects in pencil-beam surveys. We continue with our study of BP stars in the Galaxy by  designing two selecting boxes corresponding to the bright and faint samples, with $12 \leq  G_{\rm 0} \leq  14.5$ and $15.5 \leq  G_{\rm 0} \leq  18$ respectively, and with $-0.15 \leq  (G_{\rm BP}-G_{\rm RP})_{\rm 0} \leq  0.15$. We have always restricted our samples to those objects with $E(B-V) \leq  0.3$, so we avoid the presence of non-BP stars  as much as possible and, given that large-scale substructures are expected to show a gradient in the proper motion space, proper motion criteria are not considered here.

\figref{mapaBPs1} shows the density maps of BP stars in the bright and faint samples in Galactic coordinates, generated using a bin size of 1$^{\circ}$ and counting stars with RUWE $\leq$ 1.4. The high-extinction regions are observed around the plane and dominate the sky at low Galactic latitudes in the range $\ell < 150^{\circ}$. The faint sample map confirms once again that there exists a remarkable overdensity of blue stars towards the hypothetical accreted dwarf galaxy. However,  it also becomes evident that CMa is only a part of a more extended substructure of stars  in the $220^{\circ} < \ell < 300^{\circ}$ section, with colors and magnitudes compatible with that of the BP stars observed towards the core of CMa. This larger and elongated overdensity is similar to the one traced by RC stars in \cite{Bellazzini2006a} and is mainly found below the Galactic plane in those regions with lower levels of extinction.

The all-sky distribution of faint BP stars might indicate that the CMa substructure is product of the difficulties to detect  those stars in high-extinction sections in the $|b| \leq  5^{\circ}$ sky. Indeed, the core of CMa is identified in a wide area with $E(B-V) \leq 0.3$, while the extinction is higher around other tentative overdensities of BP stars found in the third Galactic quadrant.  Our density map is in good agreement with the distribution of 2MASS M-giant stars derived by \cite{Rocha-Pinto2006}, in which a stellar substructure dubbed ``Argo" is revealed. The latter overdensity seemed to be centered at $\ell \sim 290^{\circ}$ and encompassing CMa, which those authors  explained in terms of the existence of a dust extinction window towards the hypothetical accreted dwarf galaxy. Detailed molecular gas surveys covering this region also show a clear gap in its distribution along this line of sight \citep[e.g.][and references therein]{Reid2019}. Therefore, one would assume that lower levels of extinction in this region may lead us to observe a completely different morphology for CMa. A significant group of {\it Gaia} objects classified as faint BP stars according to our criteria is found towards $(\ell, b) = (65^{\circ}, 5^{\circ})$, whose extension is unknown due to the high levels of extinction in its surrounding sky area. An overdensity at this same position was also reported by \cite{Martin2004} in their CMa discovery paper. However, other substructures mentioned in that paper (e.g. the Northern Arc) are not detected in our study.

Interestingly, \cite{Carraro2017} obtained radial velocities for the stellar populations along the line of sight to the open cluster Tombaugh\,1 and included a group of stars with colors and magnitudes compatible with that of our BP stars, with $d_{\odot} > 5$\,kpc, and likely members of such a vast structure. When we cross-match their star sample C with {\it Gaia} DR2 data and computed their orbits using the same setup described in \secref{orbita},  they also seem to be confined to the plane. These authors argued that these stars are members of the Outer spiral arm of the Milky Way \citep[e.g. see ][]{Du2016,Benedettini2020}, which is displaced from the Galactic plane due to the warped disk observed in the outer Galaxy. In a similar way, CMa seems to be a section of the Outer (also Norma-Outer) spiral arm, which is observed thanks to the unlikely combination of low interstellar extinction and maximum amplitude of the  Galactic warp along that line of sight.

The distribution of bright BP stars is shown in the upper panel of \figref{mapaBPs1} and also suggests that these blue and bright stars are dispersed  throughout the $|b| < 15^{\circ}$ Galaxy instead of being exclusively found in the CMa area, which is a solid argument against the extra-Galactic origin of that structure. Stars in the bright BP sample are mainly found in the northern hemisphere of the third Galactic quadrant in the $160^{\circ} \leq \ell \leq 260^{\circ}$ area. The projected position of this overdensity and its heliocentric distance ($d_{\odot} \sim 4$\,kpc) is in good agreement with the trajectory of the Perseus spiral arm \citep[see][]{Hou2014}. Besides the Perseus and Outer spiral arms, our results in  \secref{proper} show the presence of a third structure along the same line of sight at $d_{\odot} \sim 16$\,kpc, which might be associated with the elusive  continuation of the Scutum-Centaurus and Outer Scutum-Centaurus spiral arms \citep{Dame2011,Armentrout2017}. Further exploration of this area with future {\it Gaia} data releases will help us to better understand the overall morphology of the outer Milky Way. 

The origin of the Monoceros ring, the halo substructure of which CMa has been proposed as the progenitor dwarf galaxy,  is still under debate. Recent numerical simulations have shown that if such a massive Galactic substructure is the result of an accretion event,  its progenitor is on a retrograde orbit and currently located behind the bulge \citep{Guglielmo2018}. On the other hand, Monoceros might be composed of stars originated in the Galactic disk and subsequently pushed away from the Galactic plane because of the interaction with one or several satellite galaxies \citep[e.g.][]{Xu2015,Laporte2018,Sheffield2018}. In either of these two possible scenarios, and based on the spatial and orbital properties derived above,  CMa is unlikely associated with Monoceros, independently of its formation processes.

\subsection{CMa  as a signature of the warped Milky Way disk}

\begin{figure}
     \begin{center}
      \includegraphics[scale=0.5]{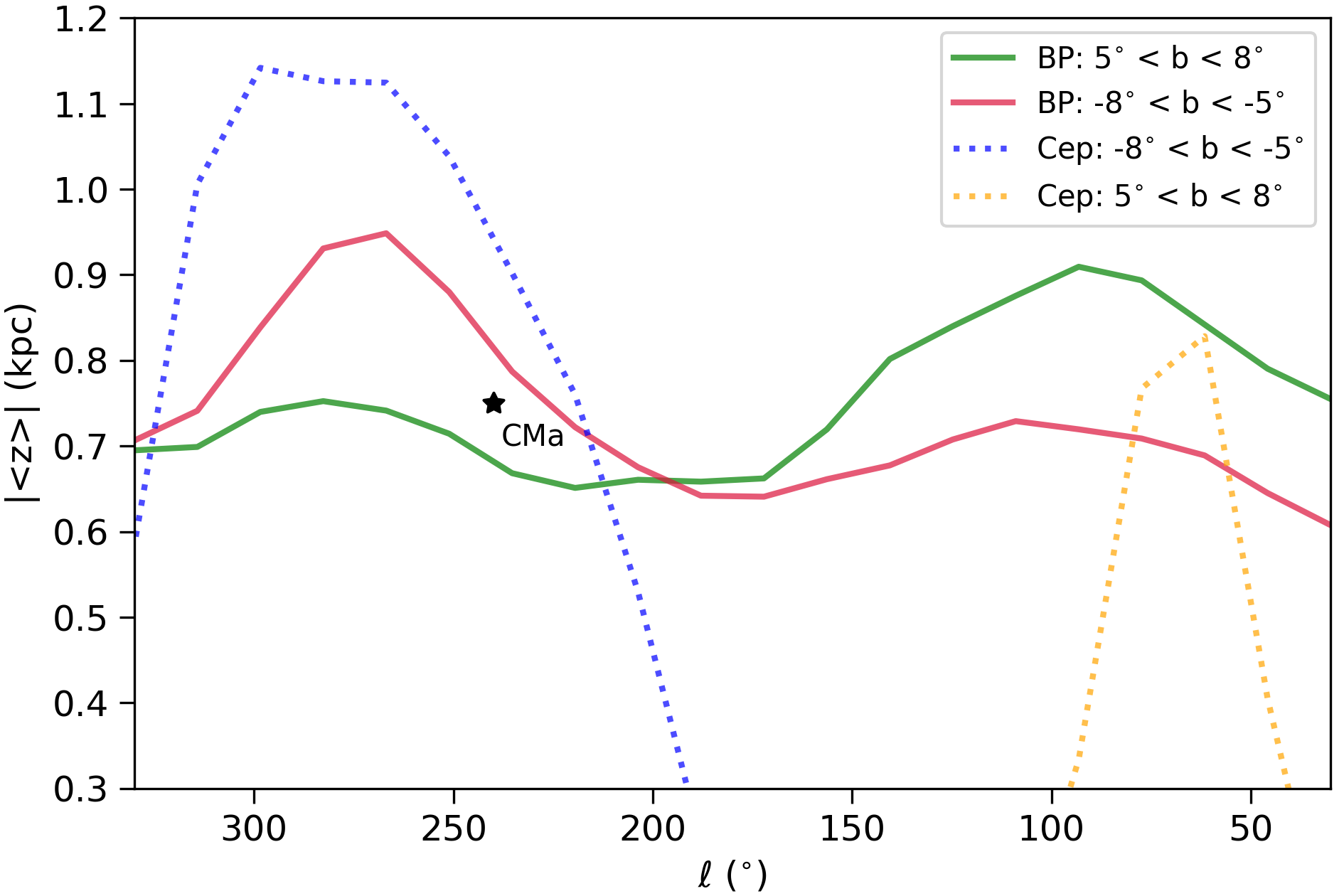}
      \caption[]{Mean value of $z$ for selected BP stars along the stripes defined by $-8^{\circ} \leq b \leq -5^{\circ}$ (red solid line) and $5^{\circ} \leq b \leq 8^{\circ}$ (green solid line). Dashed lines correspond to the projected position of the \cite{Chen2019} sample of Cepheid variable stars along the same sky stripes.} 
\label{mapa_warp}
     \end{center}
   \end{figure}

As we have mentioned above, the fact that CMa is found below the Galactic plane in the third quadrant suggests that it is likely  associated with the observational signature produced by the projected warped disk, as previously suggested by several authors \citep[e.g.][]{Momany2006,Lopez-Corredoira2007,Carraro2017}. Although the study and the parametrization of the Galatic warp is beyond the scope of this work \citep[see recent results on this topic in][]{Poggio2018,Chen2019,Romero-Gomez2019,Chrobakova2020,Poggio2020}, we explore here the vertical distribution of BP stars in the faint sample, with colors and magnitudes compatible with those of the BP stars towards CMa, as a function of their Galactic longitude. We estimate the mean $z$ values for all stars in the sky stripes defined by $-8^{\circ} \leq b \leq -5^{\circ}$ and $5^{\circ} \leq b \leq 8^{\circ}$, using the main parallax of the stars in each of the 20 bins in which we have divided the $30^{\circ} \leq \ell \leq 330^{\circ}$ area. The resulting vertical distribution for these objects is found in \figref{mapa_warp}. 

Both components of BP stars seem to be confined to the |$z$| $< 1$\,kpc section of the Milky Way and reach their maximum separation from the Galactic plane at $\ell \sim 270^{\circ}$ and $\ell \sim 90^{\circ}$, while the line of nodes is near parallel to the solar Galactocentric radius. However, as we consider the samples separately, it is evident that the |<\,$z$\,>| values are higher along the third and fourth Galactic quadrants for stars below the plane, while BP stars with positive Galactic latitudes reach |<\,$z$\,>| $\sim 0.9$\,kpc along the first and second quadrants. This behavior for BP stars is easily observed in the map shown in the lower panel of  \figref{mapaBPs1} and seem to put in evidence the warped structure of the Galactic disk, with a distortion towards positive latitudes in the $\ell < 180^{\circ}$ region and towards negative latitudes in the $\ell > 180^{\circ}$ section of the Milky Way.

In this context, CMa is located in a region of the Galaxy where the warp of the disk, as traced by BP star candidates, is closed to reach its maximum distance to the $b = 0^{\circ}$ level. In \figref{mapa_warp} we have also overplotted the sample of Cepheid stars identified by \cite{Chen2019} in the same sky stripes used here, whose projected positions in the $z - \ell$ plane are in good agreement with the warped structure revealed by our sample of BP stars. The Galactic longitude ranges in which the warped disk is observed are similar to those unveiled by \cite{Poggio2018}, although their stars with higher vertical velocities are found at $\ell > 120^{\circ}$.  These results reinforce the scenario in which the bright BP population away from the disk is mainly associated with the Galactic warp, and offers an excellent opportunity to study the formation and stellar population of this structure with unprecedented detail thanks to future {\it Gaia} data releases.

\section{Conclusions}

We have revisited the stellar population, extension and kinematic of the putative CMa overdensity with DECam and {\it Gaia} DR2 trying to shed light on its real three-dimensional structure and its origin as possible debris of an accreted dwarf galaxy. The proper motion values for a sample of BP stars, hypothetically associated with the core of CMa, are concentrated in the proper motion space but still with values compatible with those of the rest of stars in the same line of sight. The orbit derived for this overdensity from those {\it Gaia} proper motion estimates suggests that it is orbiting the Milky Way with an on-plane rotation. 

We are able to trace the CMa overdensity using BP stars selected in our photometric survey. The CMa overdensity is a real structure at low Galactic latitudes but, when we extend our study to the rest of the sky using {\it Gaia} photometry, we find that BP stars are observed not only within the central areas of the CMa overdensity but also across the third Galactic quadrant and the rest of the Galaxy.

We conclude that the CMa properties are more consistent with a structure, likely the Outer spiral arm, embedded in the warped disk, instead of being a halo substructure generated by the assimilation of a minor satellite galaxy. Future {\it Gaia} data releases will provide us accurate astrometry, photometry and radial velocities not only for the BP population across the Galaxy but also for the faint end of the CMD, where the MS of the stellar system  associated with those blue and bright stars is present. Such a remarkable dataset will allow us to understand some of the most tangled Galactic features, including the flare and warp of the Milky Way disk.

\section*{Data Availability}

This project used data obtained with the Dark Energy Camera (DECam; proposal ID 2013A-0615) and from the European Space Agency (ESA) mission {\it Gaia} (\url{https://www.cosmos.esa.int/gaia}), processed by the {\it Gaia} Data Processing and Analysis Consortium (DPAC, \url{https://www.cosmos.esa.int/web/gaia/dpac/consortium}). 

\section*{Acknowledgements}

Thanks to the anonymous referee for her/his helpful comments and suggestions. We thank to Santi Cassisi for his help to compute the synthetic color-magnitude diagram used in this study and to X. Chen for kindly providing the positions of their sample of Cepheids. JAC-B acknowledges financial support to CAS-CONICYT 17003. DMD acknowledges support from the Spanish Ministry for Science, Innovation and Universities and FEDER funds through grant AYA2016-81065-C2-2. EJA and DMD acknowledge financial support from the State Agency for Research of the Spanish MCIU through the "Centre of Excellence Severo Ochoa" award for the Instituto de Astrof\'isica de
Andaluc\'ia (SEV-2017-0709) and from grant PGC2018-095049-B-C21. DECam was constructed by the Dark Energy Survey (DES) collaboration. Funding for the DES Projects has been provided by the U.S. Department of Energy, the U.S. National Science Foundation, the Ministry of Science and Education of Spain, the Science and Technology Facilities Council of the United Kingdom, the Higher Education Funding Council for England, the National Center for Supercomputing Applications at the University of Illinois at Urbana-Champaign, the Kavli Institute of Cosmological Physics at the University of Chicago, the Center for Cosmology and Astro-Particle Physics at the Ohio State University, the Mitchell Institute for Fundamental Physics and Astronomy at Texas A\&M University, Financiadora de Estudos e Projetos, Funda{\c c}{\~a}o Carlos Chagas Filho de Amparo {\`a} Pesquisa do Estado do Rio de Janeiro, Conselho Nacional de Desenvolvimento Cient{\'i}fico e Tecnol{\'o}gico and the Minist{\'e}rio da Ci{\^e}ncia, Tecnologia e Inovac{\~a}o, the Deutsche Forschungsgemeinschaft, and the Collaborating Institutions in the Dark Energy Survey. Funding for the {\it Gaia} DPC has been provided by national institutions, in particular the institutions participating in the {\it Gaia} Multilateral Agreement.

\def\jnl@style{\it}                       
\def\mnref@jnl#1{{\jnl@style#1}}          
\def\aj{\mnref@jnl{AJ}}                   
\def\apj{\mnref@jnl{ApJ}}                 
\def\aap{\mnref@jnl{A\&A}}                
\def\apjl{\mnref@jnl{ApJL}}               
\def\mnras{\mnref@jnl{MNRAS}}             
\def\nat{\mnref@jnl{Nat.}}                
\def\iaucirc{\mnref@jnl{IAU~Circ.}}       
\def\atel{\mnref@jnl{ATel}}               
\def\iausymp{\mnref@jnl{IAU~Symp.}}       
\def\pasp{\mnref@jnl{PASP}}               
\def\araa{\mnref@jnl{ARA\&A}}             
\def\apjs{\mnref@jnl{ApJS}}               
\def\aapr{\mnref@jnl{A\&A Rev.}}          

\bibliographystyle{mn2e}
\bibliography{biblio}

\bsp	
\label{lastpage}
\end{document}